\documentstyle[aps,prl]{revtex}
\input epsf.sty
\begin{document}

\twocolumn[\hsize\textwidth\columnwidth\hsize\csname@twocolumnfalse\endcsname

\draft
\title{Random-Mass Dirac Fermions in an Imaginary 
Vector Potential: \\
Delocalization Transition and Localization Length}
\author{Koujin Takeda\cite{Takeda}}
\address{Institute for Cosmic Ray Research, University of Tokyo, Tanashi, 
Tokyo, 188-8502 Japan
}
\author{Ikuo Ichinose\cite{Ichinose}}
\address{Institute of Physics, University of Tokyo, 
Komaba, Tokyo, 153-8902 Japan 
}
%\date{, 2000}
\maketitle

\begin{abstract}
One dimensional system of Dirac fermions with a random-varying mass 
is studied by the transfer-matrix methods which we developed recently.
We investigate the  effects of nonlocal correlation of the spatial-varying 
Dirac mass on the delocalization transition.
Especially we numerically calculate both the ``typical" and 
``mean" localization
lengths as a function of energy and the correlation length of 
the random mass.
To this end we introduce an imaginary vector potential as suggested
by Hatano and Nelson and solve
the eigenvalue problem.
Numerical calculations are in good agreement with the results 
of the analytical calculations.

\end{abstract}


]

%\begin{multicols}{2}
One of the most important problems in condensed matter
physics is the localization phenomena in random-disordered 
systems\cite{Anderson}.
At present it is believed that all states tend to localize
in disorder systems in two and lower dimensions. 
In some special cases, however, some specific states remain
extended even in the presence of strong disorders.  
System of Dirac fermions with a random-varying mass in one
dimension has been studied from this point of 
view\cite{Ovc,Comtet,DSF,BF,Mathur,Shelton,Gog}.
In the previous papers\cite{TTIK,IK,IK2} we studied the 
effect of nonlocal correlation
of the random mass on the extended states which exist near the band
center.
For numerical studies, we reformulate the system by transfer-matrix 
formalism, and obtained
eigenvalues and wave functions for various configurations
of random telegraphic mass\cite{TTIK}.
We verified that the density of states obtained by the transfer-matrix 
methods is in good agreement with the analytical calculation 
in Ref.\cite{IK}.
In this paper we shall introduce an imaginary vector potential into
the system of the random-mass Dirac fermions and study a 
localization-delocalization phase transition by varying the magnitude of
the vector potential.
Through this study we shall obtain the localization length of the states as 
a function of energy and the correlation length of the random mass.
This method of calculating the localization length is based on the idea
by Hatano and Nelson\cite{HN}.
 
We shall consider a Dirac fermion in one spatial dimension with a 
coordinate-dependent mass $m(x)$ and in an imaginary vector potential $g$,
whose Hamiltonian is given by,
\begin{eqnarray}
{\cal H}&=&\int dx \psi^\dagger h\psi,\\
h&=&-i\sigma^z (\partial_x+g) +m(x)\sigma^y,
\end{eqnarray}
where $\vec{\sigma}$ are the Pauli matrices. 
This fermion model is an low-energy effective model of random-hopping 
tight binding models and random-bond spin chains\cite{BF}.
We introduce the components of $\psi$ as $\psi=(u,v)$.
In terms of them the Dirac equation is given as,
\begin{eqnarray}
\left(\ \frac{d}{dx}\ +\ g +\ m(x)\ \right) u(x)&=&Ev(x),\nonumber\\
\left(\ -\frac{d}{dx}\ -\ g +\ m(x)\ \right) v(x)&=&Eu(x). 
\label{eq:dirac1}
\end{eqnarray}
(We follow the notations in Ref.\cite{Comtet}.) 
From Eqs.(\ref{eq:dirac1}), we obtain the
Schr$\ddot{\mbox{o}}$dinger equations,
\begin{eqnarray}
 && \left(-\frac{d^2}{dx^2} - 2g \frac{d}{dx} -m'(x)+
(m^2(x)-g^2)\right) u(x)
  \nonumber \\ 
 &&\hspace{5cm} = E^2u(x),
\label{eq:schroedinger1}
\end{eqnarray}
and similarly for $v(x)$.

In this paper we restrict the shapes of $m(x)$ to
multi-soliton-antisoliton configurations\cite{NS}. 
The multi-soliton-antisoliton configurations
are given by,
\begin{eqnarray}
m(x)&=&\sum_i\bar{m}(\theta(x-\alpha_i)-1) \nonumber  \\
&&+\sum_j\bar{m}
(\theta(-x+\beta_j)-1),
\label{stepm}
\end{eqnarray}
where $\alpha_i$'s($\beta_j$'s) are positions of solitons(anti-solitons).
An example of $m(x)$ is given in Fig.1.
If we vary the distances $l$ between soliton and anti-soliton
according to the exponential distribution like,
\begin{equation}
P(l)={1 \over 2\tilde{\lambda}} \exp 
\Big(-{l\over 2\tilde{\lambda}}\Big),
\end{equation}
where $\tilde{\lambda}$ is a parameter, then 
$m(x)$ has the following correlation\cite{Comtet},
 \begin{equation}
 [\ m(x)\ m(y)\ ]_{{\rm ens}} = \frac{A}{\tilde{\lambda}} 
 \exp\ (-|x-y| / \tilde{\lambda}),
 \label{disordercor}
 \end{equation} 
where $\sqrt{\frac{A}{\tilde{\lambda}}}$ essentially corresponds to 
the height of the soliton and 
anti-soliton, i.e., $\bar{m}$ in (\ref{stepm}).
From (\ref{disordercor}), $\tilde{\lambda}$ is the correlation length
of the random mass and the limit $\tilde{\lambda} \rightarrow 0$ corresponds
to the white-noise case.
In subsequent papers, we shall study Dirac fermions
with long-range correlated random mass by using the methods
examined in this paper\cite{TI}.
There we expect some interesting phenomena like existence
of nontrivial mobility edge, nonuniversality of the multi-fractal
exponents, etc.
Studies in this paper show that the imaginary-vector-potential methods
for calculating the localization lengths are reliable and we shall
use them for studies on random systems with long-range correlated
disorder.
\begin{figure}
\label{fig:figure1}
\begin{center}
\unitlength=1cm

\begin{picture}(15,4)
\centerline{
\epsfysize=5cm
\vspace{0mm}
\epsfbox{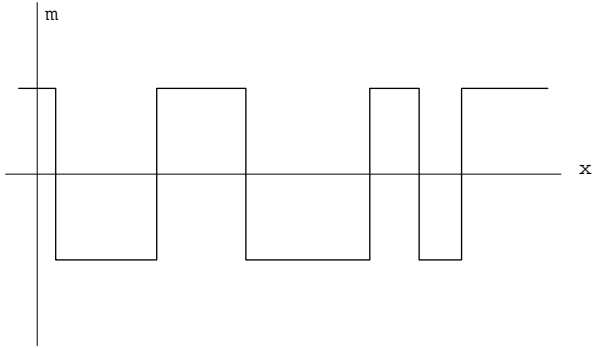}
}
\end{picture}
\vspace{0mm}
\caption{An example of configurations of solitons and anti-solitons.}

\end{center}
\end{figure}

For the vanishing imaginary vector potential,
we solved the Schr$\ddot{\mbox{o}}$dinger equations in (\ref{eq:schroedinger1})
under the periodic boundary condition with various multi-soliton-antisoliton 
configurations of $m(x)$ using the transfer-matrix method, 
and obtained the energy spectrum and wave functions\cite{TTIK}. 

Effect of the imaginary vector potential was discussed by Hatano and 
Nelson\cite{HN}. 
Let us denote the eigenfunction of energy $E$ for $g=0$ as $\Psi_{0}(x)$, 
and suppose the shape of $\Psi_{0}(x)$ as
\begin{equation}
 \Psi_{0}(x) \cong \exp \left(  -\frac{|x-x_{c}|}{\xi_{0}} \right),
\end{equation}    
where $\xi_{0}$ is the localization length
and $x_{c}$ is the center of this localized state.
When we turn on the constant imaginary vector potential $g$, 
the eigenfunction is obtained from
$\Psi_{0}(x)$ by the ``imaginary" gauge transformation,
\begin{equation}
 \Psi(x) \cong \exp \left( -\frac{|x-x_{c}|}{\xi_{0}} -\ g\ (x-x_{c})
 \right). 
\label{psix}  
\end{equation}
This means that the localization length of this eigenfunction is
\begin{eqnarray}
 \xi_{g} & = &  \frac{\xi_{0}}{1+g\xi_{0}} \ ( x>x_{c} ) \nonumber \\
 \xi_{g} & = &  \frac{\xi_{0}}{1-g\xi_{0}} \ ( x<x_{c} ).
\end{eqnarray}
At the point $g=1/\xi_{0}$, localization length for $x<x_{c}$ diverges, 
and if the imaginary vector potential $g$ is increased more, 
$\xi_{g}$ for $x<x_{c}$ becomes negative, 
and this eigenfunction cannot satisfy the periodic boundary condition
even if the length of system is large enough. 
So if localized eigenstate with energy $E$ disappears at $g=g_{c}$ as $g$ is
increased, then the localization length of the eigenstate $\Psi_{0}(x)$
is $1/g_{c}$.
Actually it is shown that for $g>g_c$, energy eigenvalue of the state
has an imaginary part and the state is extended as we shall see
shortly\cite{HN}.

The transfer-matrix methods can be easily extended for the case
of nonvanishing $g$.
We obtain energy eigenvalues and eigenfunctions numerically.
First of all, we show that the delocalization transition actually
occurs.
In Fig.2, we show the wave functions of a low-lying state
in vanishing and nonvanishing imaginary vector potential.
For $g=0$, the state is obviously localized whereas at $g=0.03$
the state becomes extended and the energy eigenvalue has an imaginary
part.
As discussed in Ref.\cite{HN}, the density distribution of
a particle is given by $|\Psi(x,-g)\Psi(x,g)|$ where $\Psi(x,-g)$
is equal to the left eigenfunction.
\begin{figure}
\label{fig:figure2}
\begin{center}
\unitlength=1cm

\begin{picture}(15,4)
\centerline{
\epsfysize=5cm
\epsfbox{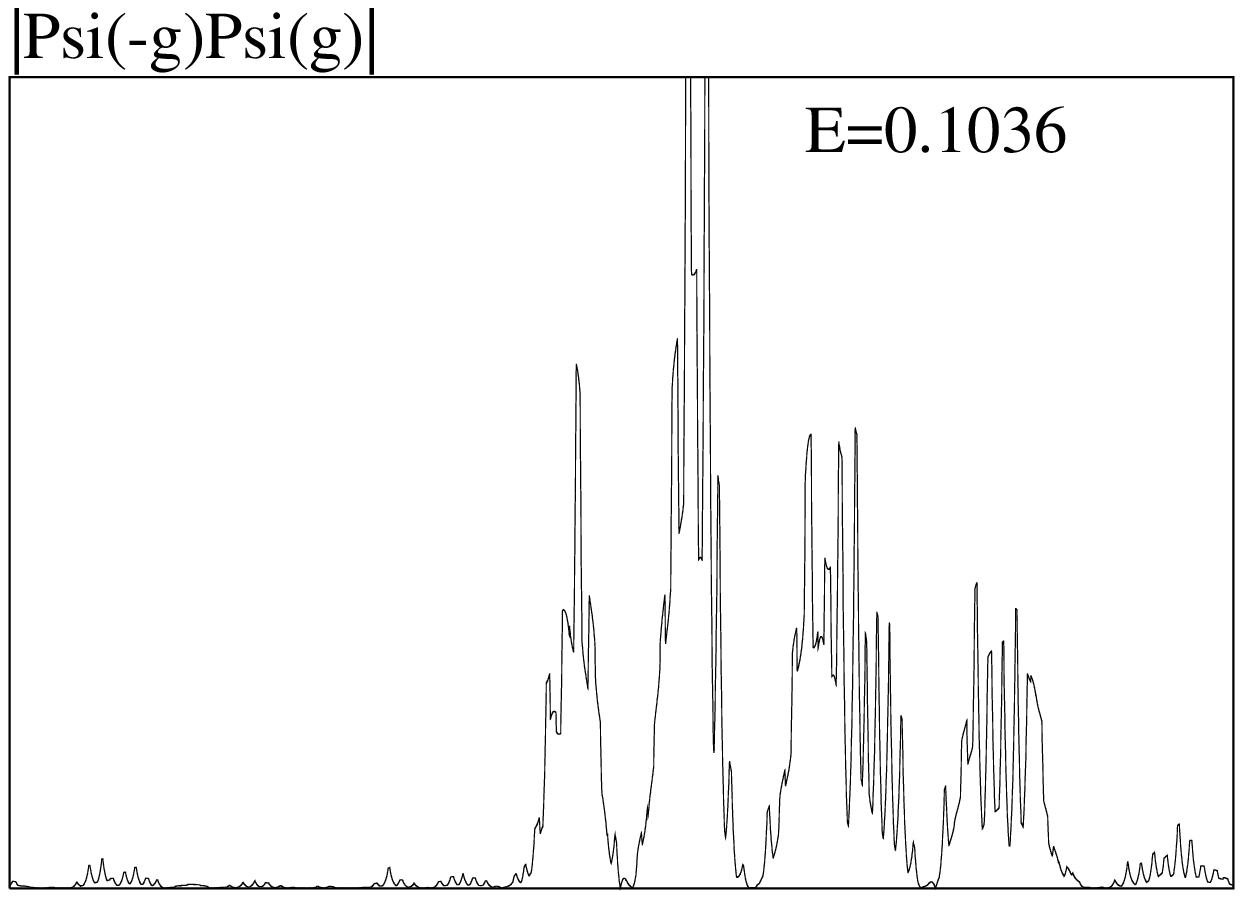}
}
\end{picture}

\begin{picture}(15,5)
\centerline{
\epsfysize=5cm
\epsfbox{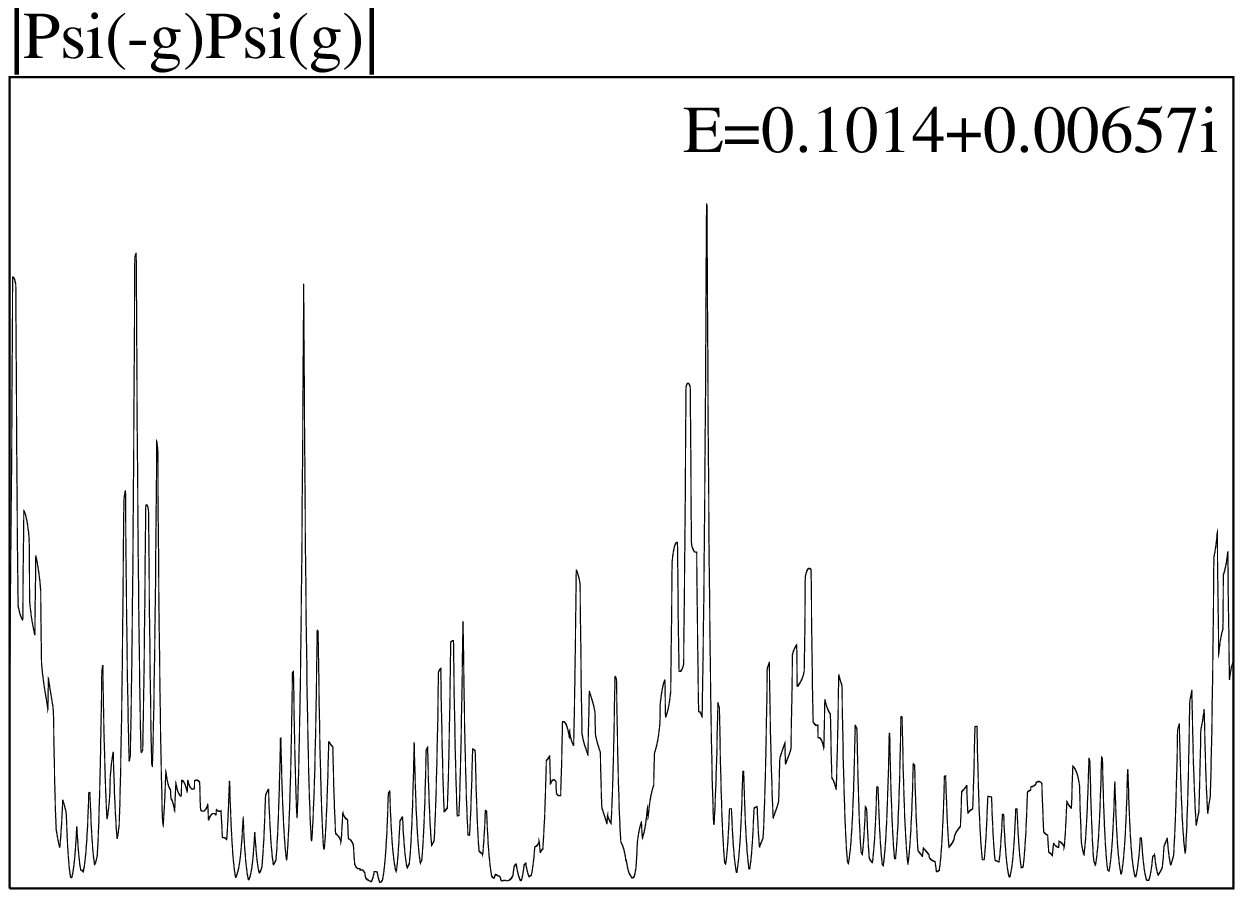}
}
\end{picture}
\vspace{5mm}
\caption{An example of the localized and the extended wave functions
in vanishing and nonvanishing imaginary vector potential $g=0$ and $g=0.03$.
The state of $g=0.03$ has a complex energy eigenvalue.}

\end{center}
\end{figure}

Let us turn to the localization length.
For the white-noise case $ [m(x)m(y)]_{\it{ens}} = A\ \delta(x-y)$, 
``typical" localization length or the inverse of the Lyapunov exponent
was obtained as\cite{Comtet,Dyson}
\begin{equation}
 \xi_t(E) = |\ln E/2A|.
\label{eq:localize1} 
\end{equation}
Numerically the typical localization length $\xi_t(E)$ is obtained by
averaging over localization lengths of {\em all} eigenstates with energy $E$. 
On the other hand,
Balents and Fisher calculated the averaged Green function 
and obtained the mean
localization length from the spatial decay of the Green function\cite{BF}.
The result is
\begin{equation}
 \xi_m(E) = |\ln E/2A|^{2}.
\label{eq:localize2}
\end{equation}
Case of nonlocally correlated random mass was studied in 
Refs.\cite{IK,IK2} and $\xi_m(E)$ is obtained as a function of
$\tilde{\lambda}$ in Eq.(\ref{disordercor}).

By numerical calculation we obtain both the typical and mean
localization lengths.
As we mentioned above, the typical localization length is the average 
over all solutions of the Schr$\ddot{\mbox{o}}$dinger
equation ({\ref{eq:schroedinger1}),
whereas the mean localization length is determined by the states
which make dominant contributions to the Green function, i.e., which
have large localization length. 

The result of the numerical calculation of the typical
localization length of the white-noise case is given in Fig.3.  
We show the ratio of the numerical results to the analytical calculation
in Eq.(\ref{eq:localize1}) in order to compare these two results. 
Therefore if the energy dependences of the localization
lengths obtained numerically and analytically are the same, 
this ratio should be constant. 
In Fig.3, the ratio seems constant over the whole range of $E$.   

\begin{figure}
\label{fig:figure3}
\begin{center}
\unitlength=1cm

\begin{picture}(15,4)
\centerline{
\epsfysize=5cm
\vspace{0mm}
\epsfbox{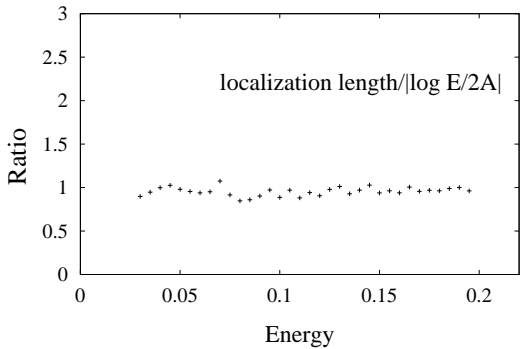}
}
\end{picture}
\vspace{0mm}
\caption{The comparison between the analytical and numerical
results of the typical localization length for the (almost) white-noise case,
i.e., small-$\tilde{\lambda}$ case.
In the numerical calculation, we set $L$(system size)$=50$, 
$\tilde{\lambda}=1/60$,
and the energy slice $\delta E=0.02$. 
This result is averaged over 2500 trials.
The analytical and numerical results are in good agreement.}

\end{center}
\end{figure}

 In Fig.4, we show the numerical results
 of the ``mean" localization length. 
 Here we use the solutions to the Schr$\ddot{\mbox{o}}$dinger equation
 which have long localization length. 
 More precisely, ``large"
 localization length $\xi$ means the one which satisfies   
 $\xi  > $ (the ``typical" localization length)+ 1.5 $\sigma$ 
 in each energy slice.
 From Fig.4, we can conclude that the energy dependence of the
 mean localization length obtained numerically is in agreement with 
 Eq.(\ref{eq:localize2}).  

\begin{figure}
\label{fig:figure4}
\begin{center}
\unitlength=1cm
 
\begin{picture}(15,4)
\centerline{
\epsfysize=5cm
\epsfbox{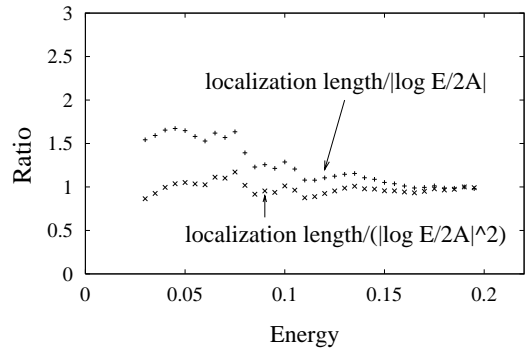}
}
\end{picture}
\vspace{0mm}
\caption{The comparison between the analytical and numerical
results of the mean localization length.
We use the data of the states which have large localization
length. (See the text.) 
We show the ratios of the numerical calculations both to 
Eqs.(\ref{eq:localize1}) and (\ref{eq:localize2}).
The ratios are normalized at $E=0.19$.}
\end{center}
\end{figure}
From the above studies, we can conclude that the above methods of calculating 
the localization lengths are reliable.

We shall turn to the case of the nonlocally-correlated disorder. 
In the white-noise limit ($\tilde{\lambda}=0$ case in Eq.(\ref{disordercor})), 
the localization lengths 
diverge only at $E=0$, that is, extended states exist only at $E=0$. 
If we let $\tilde{\lambda} > 0 $, the random mass becomes
nonlocally-correlated, and the critical energy or the mobility edge
at which the delocalization transition 
occurs may change.

We investigate the ``typical" and ``mean" localization lengths 
in the case of nonvanishing $\tilde{\lambda}$'s. 
The behaviour of the ``typical" and ``mean"
localization lengths obtained as in the white-noise case are given 
in Figs.5 and 6.
It seems that there is {\em no} $\tilde{\lambda}$-dependence
in the typical localization length.
On the other hand, Fig.6 shows that the mean localization length has 
a small but finite dependence on $\tilde{\lambda}$.
 
\begin{figure}
\label{fig:figure5}
\begin{center}
\unitlength=1cm

\begin{picture}(15,4)
\centerline{
\epsfysize=5cm
\epsfbox{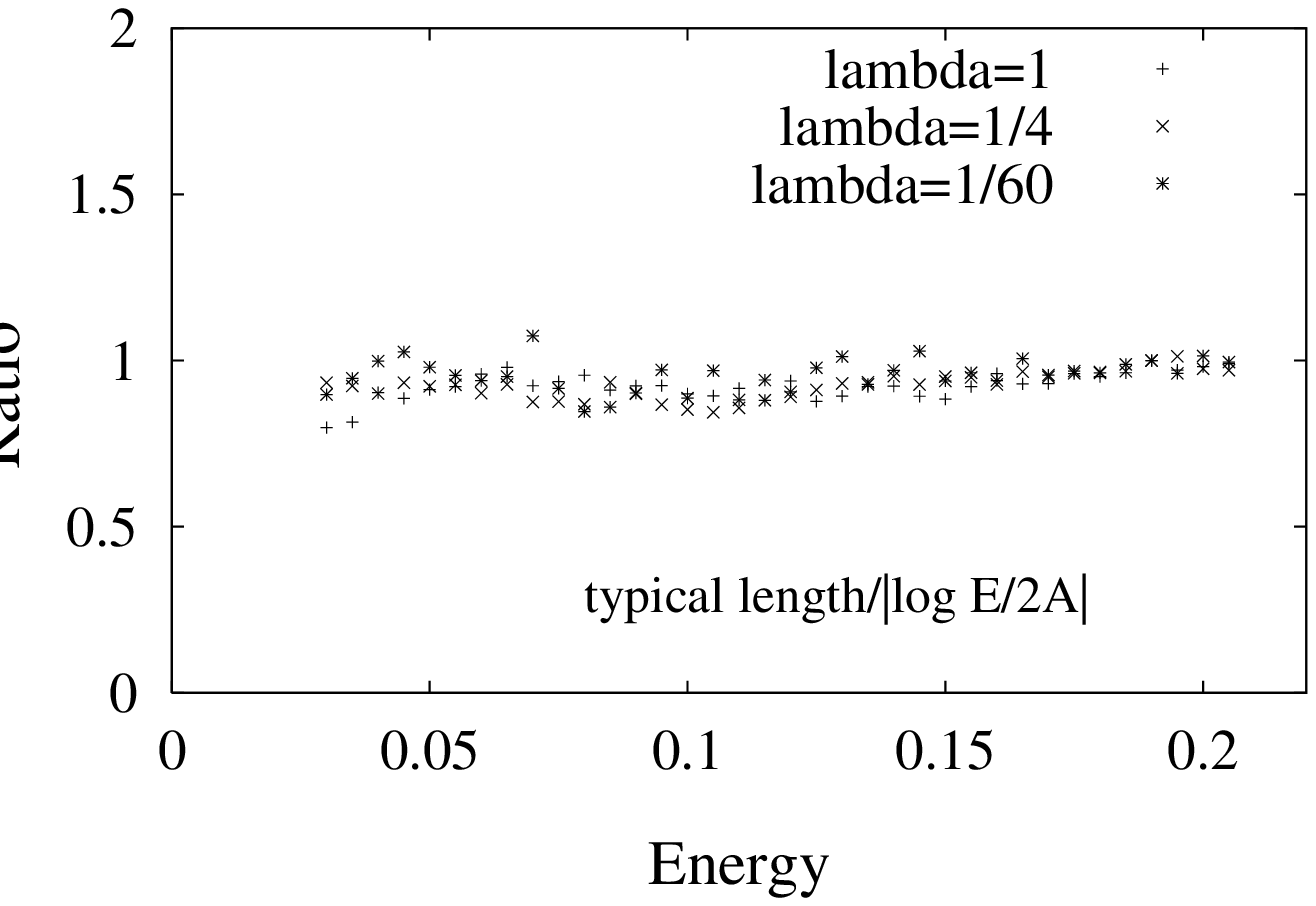}
}
\end{picture}
\vspace{0mm}
\caption{The behaviour of the localization length in the case of nonvanishing 
 $\tilde{\lambda}$'s. 
 The ratio is the ``typical" localization length to $|\ln E|$.
 The ratio is normalized at $E=0.19$.}
\end{center}
\end{figure}
 
\begin{figure}
\label{fig:figure6}
\begin{center}
\unitlength=1cm

\begin{picture}(15,4)
\centerline{
\epsfysize=5cm
\epsfbox{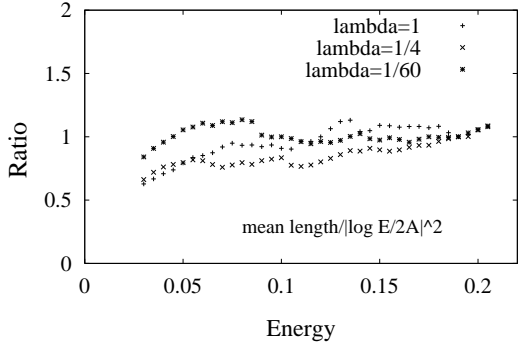}
}
\end{picture}
\vspace{0mm}
\caption{The behaviour of the localization length in the case of nonvanishing 
 $\tilde{\lambda}$'s. 
 The ratio is the ``mean" localization length to $|\ln E|^2$.
 The energy slice $\delta E=0.03$, and the ratio is normalized at $E=0.19$.}
\end{center}
\end{figure}

From the above calculations, we conclude that the effect of the 
short-range correlations in disorders is 
not so large. 
Especially the result indicates that the delocalization transition 
occurs at $E<0.03$. (If the mobility edge exists at $E_c>0$, 
the ratio in Fig.5 or 6 must diverge at $E_c$.) 
The delocalization transition probably occurs at $E=0$. 

In the previous paper\cite{IK2} we calculated the localization length 
for the random mass with the short-range correlation (\ref{disordercor}).
We obtained the ``mean'' localization length to the 1st order of 
$\tilde{\lambda}$ by means of the Green function method.
 The result is
\begin{equation}
  \xi(E) = \frac{1}{A} \Big( \frac{\ln |\frac{E}{2A}|^{2}}{\pi^{2}}
+A \tilde{\lambda} \frac{4|\ln \frac{E}{2A}|}{\pi^{2}}\Big)
+O(\tilde{\lambda}^2).
\end{equation}
In Fig.7 we show the ratios of the numerical result to the analytical
calculation up to the 0th and the 1st order of $\tilde{\lambda}$.
This shows that the analytical result with the 1st order correction 
of $\tilde{\lambda}$ is in better agreement with the numerical result, 
but the correction is small.

\begin{figure}
\label{fig:figure7}
\begin{center}
\unitlength=1cm

\begin{picture}(15,4)
\centerline{
\epsfysize=5cm
\epsfbox{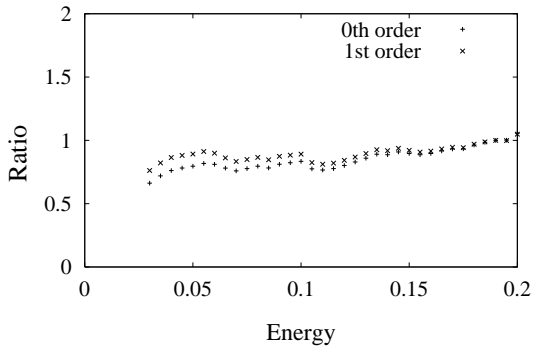}
}
\end{picture}
\vspace{0mm}
\caption{The comparison between the numerical result and the analytical one
 in the case of relatively large $\tilde{\lambda}$. 
 We show the ratio of the localization length calculated numerically 
 to the ones obtained analytically. Here we used the analytical results
 of the 0th and the 1st order of $\tilde{\lambda}$. 
 Here we set $\tilde{\lambda}=1/4$ and $L=50$.
 The energy slice $\delta E=0.03$, and the ratio is normalized at $E=0.19$.}
\end{center}
\end{figure}

From the investigations given as far we can conclude that
the numerical methods used in this paper are reliable for
calculating the localization lengths.
It is very interesting to study the case of disorders with a long-range
correlation.  
We expect that a nontrivial mobility edge $E_c>0$ exists 
for a certain long-range correlated random mass.
Actually the one-dimensional Anderson model with
long-range correlated disorder was studied\cite{Lyra}, 
and it is shown that there exists a nontrivial mobility edge.

We can also calculate exponents of the multi-fractal 
scaling\cite{Huckenstein} by the transfer-matrix methods\cite{TTIK}.
These problems are under study and results will be reported
in a future publication\cite{TI}.

%\end{multicols}

\end{document}